# Digital biomarkers and artificial intelligence for mass diagnosis of atrial fibrillation in a population sample at risk of sleep disordered breathing

Armand Chocron[1,2], Roi Efraim[3], Franck Mandel[4], Michael Rueschman[5], Niclas Palmius[6], Thomas Penzel[7,8], Meyer Elbaz[4] and Joachim A. Behar[1]

[1] Faculty of Biomedical Engineering, Technion, Israel Institute of Technology, Haifa, Israel

[2] Faculty of Electrical Engineering, Technion, Israel Institute of Technology, Haifa, Israel

[3] Cardiology department, Rambam Hospital, Haifa, Israel

[4] CHU Rangueil, Toulouse, France

[5] Program in Sleep Medicine Epidemiology, Division of Sleep and Circadian Disorders, Department of Medicine, Brigham and Women's Hospital, Boston, MA

[6] Wolfson College, Oxford OX2 6UD, UK

[7] Interdisciplinary Center of Sleep Medicine, Charité Universitätsmedizin Berlin, Germany

[8] Saratov State University, Saratov, Russia

**Corresponding author:**

Joachim A. Behar

Faculty of Biomedical Engineering

Technion-IIT, Haifa, Israel

jbehar@technion.ac.il

**Abstract**

**Study Objectives:** Atrial fibrillation (AF) is the most prevalent arrhythmia and is associated with a five-fold increase in stroke risk. Many individuals with AF go undetected and thus untreated. These individuals are often asymptomatic or have paroxysmal AF. There are ongoing debates on whether mass screening for AF in the general population is to be recommended. However, there is incentive in performing systematic screening for specific at risk groups such as individuals suspected of sleep-disordered breathing where an important association between AF and obstructive sleep apnea (OSA) has been demonstrated. In this research we introduce a new methodology leveraging digital biomarkers and recent advances in artificial intelligence (AI) for the purpose of mass AF diagnosis from hours of single channel electrocardiogram (ECG) recording. We demonstrate the value of such methodology in a large population sample at risk of sleep disordered breathing.

**Methods:** Four databases, totaling n=3,088 patients and p=26,913 hours of continuous single channel electrocardiogram raw data were used. Three of the databases (n=125, p=2,513) were used for training a machine learning model in recognizing AF events from beat-to-beat interval time series. The visit 1 of the sleep heart health study database (SHHS1, n=2,963, p=24,400) consists of overnight polysomnographic (PSG) recordings, and was considered as the test set to evaluate the feasibility of identifying prominent AF rhythm from PSG recordings. In SHHS1, expert inspection identified a total of 70 patients with a prominent AF rhythm.

**Results:** Model prediction on the SHHS1 showed an overall $Se = 0.97$, $Sp = 0.99$, $NPV = 0.99$, $PPV = 0.67$ in classifying individuals with or without prominent AF. $PPV$ was non-inferior (p=0.03) for individuals with an apnea-hypopnea index (AHI) $\geq$ 15 versus AHI < 15. Over 22% of correctly identified prominent AF rhythm cases were not documented as AF in the SHHS1.

**Conclusions:** Individuals with prominent AF can be automatically diagnosed from an overnight single channel ECG recording, with an accuracy unaffected by the presence of moderate to severe OSA. AF detection from overnight single channel ECG recording revealed a large proportion of undiagnosed AF and may enhance the phenotyping of OSA by identifying individuals in whom cardiac function may have been affected by OSA.

# 1. Introduction

*Atrial fibrillation*

Atrial fibrillation (AF) is the most common arrhythmia, with a prevalence of 0.95% among the adult population [1]. It is associated with quivering or irregular heartbeat, that can lead to blood clots, stroke, heart failure and other heart-related complications [2]. There exist treatments for AF including cardioversion and cardiac ablation as well as drugs intending at controlling the heart rate [3]. The currently accepted convention for AF diagnosis is the presence of an episode lasting at least 30 seconds [3]. Many individuals with AF go undetected and thus untreated because these are often asymptomatic or have paroxysmal AF (PAF) i.e. episodes of AF that occur occasionally. Yet, there are ongoing debates in the medical community as per whether mass screening for AF in the general population is to be recommended because of the costs involved and uncertainty over the benefits [4]. However, there may be incentive in performing such systematic screening for specific at high-risk groups [4].

*Value for systematic AF screening during sleep*

We motivate the clinical relevance and feasibility for performing overnight systematic AF screening in population sample at risk of sleep disordered breathing[5] with three arguments: (1) *Interaction of AF with sleep disordered breathing*: there is incentive in performing systematic AF screening in individuals suspected of sleep-disordered breathing. Several pivotal studies conducted over the past decade have highlighted a strong association between obstructive sleep apnea (OSA) and AF [6]. For example it was shown that individuals with severe sleep-disordered breathing have fourfold higher odds of AF than those without sleep-disordered breathing after adjustment for potential confounders [7]. Given the high prevalence of AF[1] and OSA[8] and the relationship that exists between AF and OSA, there is a high motivation for mass AF screening from sleep studies. Yet, to date, very little research has been done regarding the analysis of AF during sleep. This is despite the fact that we spend about one third of our lives sleeping. (2) *Longer recordings:* short recordings will lead to missing PAF individuals. In addition, it has been shown that further characterization of the condition over longer intervals, lasting from hours to days, may improve phenotyping of the diseases[9,10]. This motivates performing long continuous electrophysiological recording for the purpose of mass AF screening. Performing an overnight recording will typically enable to collect 6-7 hours of continuous data. (3) *Diseases expression during sleep*: previous research has suggested that there exists diurnal variation in the timing of paroxysmal atrial fibrillation (PAF) events, with peak incidence at nighttime [11,12]. This further motivates leveraging sleep recordings for the purpose of AF mass screening.

*Digital health and novelty of the approach taken*

The creation of intelligent algorithms combined with existing and novel wearable biosensors offer an unprecedented opportunity to improve the diagnosis and monitoring of patients with AF. We present a paradigm per which we intend to use continuous single ECG channel overnight recordings for the purpose of mass AF screening in a population at risk of sleep disordered breathing. We present this paradigm as novel when comparing with the most recent research attempting at performing mass AF screening from video camera [13] or smartwatches pulse recording [14,15]. Indeed, video camera usage has the intrinsic limitation of being limited in time since the user must hold his finger in front of the phone camera. This will lead to numerous undetected PAF cases and the inability to further characterize AF (e.g. by providing the AF burden). Smartwatches usage such as in the Apple Heart Study have the advantage of enabling longer continuous recordings and analysis. However, in their methodology only intermittent spot tachograms were recorded and in the case an irregular pulse was detected, then the algorithm would prospectively and opportunistically scan for more irregularity during minimal arm movement. This was likely performed to reduce false positives due to noise induced movement in the photoplethysmography signal.

*Working hypothesis*

In this work, we hypothesize that nocturnal AF events can be automatically and accurately identified using a machine learning (ML) approach applied to a single channel ECG recording. We further hypothesize that the positive predictive value will not be impaired by the presence of OSA. Such proof will pave the way to the creation of novel digital health solutions for systematic AF screening in portable sleep studies.

## 2. Methods

### 2.1 Databases

A total of four databases [16,17] totaling n=3,088 patients and p=26,913 hours of data were used.

*Training set databases*

*The PhysioNet MIT-BIH Normal Sinus Rhythm (NSR) database:* The NSR database [17] is composed of n=18 long-term, 21.2 ± 1.2 hours long, electrocardiogram (ECG) recordings, totaling p=384 hours. The original ECG measurements were sampled at 128 Hz. The subjects had been referred to the arrhythmia laboratory at Boston's Beth Israel Hospital (BIH) and had no significant arrhythmias apart from the presence of ectopic beats. Subjects included 5 men, aged 26-45 years, and 13 women, aged 20-50 years.

*The PhysioNet Long-Term AF (LTAF) database*: The LTAF database[17] consists of recordings of n=84 individuals suffering from PAF or sustained AF. Each record contains two simultaneously recorded ECG signals digitized at 128 Hz, with 12-bit resolution over a 20 mV range; record durations are of 22.7 ± 2.4 hours long. The overall database totals p=1,900 hours, including 874 hours in AF and 1,026 hours spent in non-AF rhythms. The original recordings were digitized and automatically annotated at Boston's Beth Israel Deaconess Medical Center. R-peak and rhythms annotation were available. These were obtained by manual review of the output of an automated ECG analysis system (PocketECG system, MEDICALgorithmics', Warsaw, Poland).

*Test set databases*

*The PhysioNet MIT-BIH AF (BIHAF) database:* This database[17] includes n=25 long-term ECG recordings of human subjects with AF (mostly PAF), totaling p=229 hours of data, with 92 hours in AF and 137 hours in non-AF rhythms. Of these, 23 records included two ECG channels, sampled at 250 Hz, with 12-bit resolution, over a range of ±10 millivolts. Two recordings did not have the raw ECG signal available and were excluded from the present analysis. Reference R-peak and rhythms annotation were available.

*The Sleep Heart Health Study (SHHS)* [16,18–20] is a multi-center cohort study implemented by the National Heart Lung & Blood Institute (ClinicalTrials.gov Identifier: NCT0000527) to determine the cardiovascular and other consequences of sleep-disordered breathing. In all, 6,441 men and women aged 40 years and older were enrolled between November 1, 1995 and January 31, 1998 to take part in SHHS Visit 1 (SHHS1). During exam cycle 3 (January 2001-June 2003), a second PSG was performed during the SHHS Visit 2 (SHHS2) for 3,295 of the participants. No distinction was made between AF and atrial flutter (AFL) and thus all individuals with documented AF or AFL were labelled "AF". The original ECG used for the purpose of AF diagnosis were obtained from standard resting 12-flead ECG, collected with the participant in a supine position, from parent cohorts. ECGs were sampled at 128 Hz and recorded for ten seconds for all leads (I, II, III, aVR, aVL, aVF, V1-V6) using a Marquette MAC PC or MAC II system. Among participants that had PSG raw data available in SHHS1, n=2,963 had AF labels. Summary statistics for age and OSA diagnosis variables are available in Table 1. However, it is not clear whether some patients originally diagnosed with AF in one of the parent studies were treated using digoxin and/or anti-coagulation (which were the options available at the time of these studies) between the time of the 12-lead ECG and the time of the PSG recordings. Thus from an epidemiological perspective, by detecting AF in the PSG recordings, we actually seek to identify AF patients whose condition manifested as an AF rhythm during SHHS1 and, thus, excluding AF patients that may have been effectively treated. Another important intrinsic limitation is that this diagnosis was based on a 10-second ECG strip, which might have missed many PAF cases. Conversely, although the Minnesota code was used to indicate persistent AF and AFL in the SHHS1 cohort, it is possible that some individuals flagged as persistent AF are actually PAF. Such mismatch would be explained by the short recording time for diagnosis that would have been performed during an AF event in a patient presenting PAF.

*Recordings re-annotation*

To account for the intrinsic limitation of the labels provided in the SHHS1 database, two intern cardiologists (co-author RE and FM) re-examined all 66 ECGs from individuals originally labelled as AF as documented in SHHS1. In addition, all recordings predicted with a high AF burden (AFB), defined by the percentage of windows predicted as AF, but not originally labelled as AF in the SHHS1, were reviewed independently by the two interns. The two interns were asked to classify a recording as AF, non-AF or low quality. Within this context, the low quality category was defined as the inability for the medical reviewer to make a decision on the presence of AF. In case of disagreement a third senior cardiologist (co-author ME), with 25 years of clinical experience, adjudicated. The review was performed using the open source software *PhysioZoo* [21], using a similar setting to that of a traditional Holter review.

## 2.2 Data processing

*Signal quality evaluation*

In order to automatically assess the quality of the raw ECG files and discard those that were noisy, we included a signal quality preprocessing step. For each 60-beat window, the *bsqi* index [22] was computed. The *bsqi* index compares the R-peaks detected by two different R-peak detectors: one reference set, usually coming from a stronger R-peak detector, and one test set, coming from another, usually weaker R-peak detector. If the two detectors agree (detect the same beats), then the quality can be assumed to be sufficiently high to reliably use the beat-to-beat time series. For the SHHS1, we used the *epltd* [23] as the reference R-peak annotations and *xqrs* [17] as the test R-peak annotations. We verified that the generated annotations contained at least 1,000 R-peak and excluded files which did not satisfy this criterion i.e. corresponding to recordings with a flat ECG. Among the remaining recordings, windows with a *bsqi* lower than 0.8 [24] were excluded from the analysis. Recordings showing a rate of exclusion, i.e., the ratio between the number of excluded windows and the total number of windows, higher than 75% were considered as corrupted and were not considered for analysis.

*HRV-based feature engineering*

A total of nine features, extracted from the statistics of the RR interval time series, were computed for each 60-beat window. Five of these features were derived from the work of Lake et al.[25], Sarkar et al.[26]. The other four features were *bsqi,* signal quality index of the window; AVNN, the mean RR interval duration; minRR, the minimal RR interval duration; and medHR, the median heart rate. A reference table including the definition of each feature is available in Table 2.

*Machine learning*

To detect AF events, we trained a random forest (RF) model (Supplementary Figure 1). The following hyperparameters were optimized using 5-fold cross-validation on the training set databases: the number of estimators and the maximal depth of the trees. The selected model included 20 different estimators with maximal depth of three. The model was trained on the nine features listed in Table 2. The model then returned a label (AF or non-AF) for each 60-beat window. The model was trained on the LTAF and NSR databases and tested on the BIHAF database. The hyperparameters of the model being optimized, the global model was generated from the three databases (LTAF, NSR and BIHAF) and applied to the SHHS1 database. The algorithm was implemented in python using the *scikit-learn* package. The method used to detect AF episodes on these databases is summarized in Figure 1.

*Classes and performance statistics*

The following statistics were computed: sensitivity ($Se$), specificity ($Sp$), positive predictive value ($PPV$), area under receiver operating characteristic curve (AUROC) and the harmonic mean between the $Se$ and $PPV$, termed the $F_1$ measure. In the context of the statistics reported for the SHHS1, these parameters are defined as: $Se$, the percentage of individuals with AF that have been correctly identified as AF out of the entire AF population; $Sp$, the percentage of individuals without AF that have been correctly identified as such, out of the whole non-AF population; $PPV$, the percentage of individuals correctly identified as having AF out of all the individuals that were predicted as AF. We used the statistical non-inferiority test recommended in Tuned da Silva et al. [28] to evaluate non-inferiority for a binary endpoint. This test was used to demonstrate non-inferiority for the model $PPV$ for the OSA versus non-OSA group. Within our context we defined population presenting an apnea hypopnea index (AHI) exceeding or equal to 15 as OSA and the population presenting an AHI lower than 15 as non-OSA. A tolerance level at 3% was used.

## 3. Results

*Signal quality*

Overall, 67 out of 2,963 patients were excluded from the SHHS1 database after the first signal quality step (Figure 2). Among the remaining recordings, 10.2% of the windows were excluded by this preprocessing step.

*Machine learning model*

Table *3* displays the results obtained by the model on the training (LTAF, NSR) and test (BIHAF) sets, and presents the results of the global model trained on three databases (LTAF, NSR and BIHAF). Performance of the RF model on the test set was $Se = 0.95$ and $Sp = 0.98$. The ROC curve for the global model is presented in Figure 3. The standard deviation in estimating the AFB on the global model was 17.6%. We took a conservative threshold at 20%, and classified, for model evaluation on the SHHS1, an individual with an estimated AFB≥20% as suffering from prominent AF. Conversely, an individual with an estimated AFB <20% was considered as a non-prominent AF patient. This threshold that is close to one standard deviation, is used as an estimation of the confidence on the algorithm performance in estimating the AFB. Taking a threshold that is low would results in a significant number of FP and impair the merit of this approach.

*Re-annotation of SHHS1 AF labels*

Out of 118 reannotated recordings (66 originally labeled as AF and 52 predicted positive by the RF model), annotator A and B agreed on 106. Among these there were a total of 65 AF cases, 39 non-AF cases and 2 cases labeled as low quality. The two low quality recordings were excluded. For the remaining 12 recordings where the two annotators disagreed, annotator A identified 9 AF, 3 non-AF and annotator B identified 3 cases as AF, 7 as non-AF and 2 as low quality. Adjudication on the 12 cases was performed by a senior cardiologist. Out of 12 cases, 5 were adjudicated as AF and 3 as non-AF. The remaining 4 recordings were too noisy for the senior cardiologist to provide a decision. These were removed from the analysis. The final labels for the annotated recordings consisted of 70 cases of AF, 42 cases of non-AF and 6 excluded files. A summary diagram is provided in Figure 2.

*Performance evaluation on the SHHS1*

Table 4 summarizes the statistics obtained on the SHHS1 for the detection of prominent AF, for both OSA (AHI≥15) and the non-OSA (AHI<15) groups based on the re-annotated labels. 103 patients (3.5%) were identified with prominent AF. The median (Q1-Q3) age for the prominent versus non-prominent AF groups were 77 (74.5-80) and 67 (53-76) years respectively (p<0.05). There were a total of 64 males (62%) in the prominent AF group against 1241 (44.5%) in the non-prominent group (p<0.05). The overall classification performance was $Se = 0.97$, and $Sp = 0.99$. For non-OSA individuals, the classification performance was $Se = 0.96$ and $Sp = 0.99$. For individuals with OSA, the classification performance was $Se = 0.98$ and $Sp = 0.99$. Non-inferiority testing for the model $PPV$ for the OSA versus non-OSA group was demonstrated (p=0.03). Figure 4 shows the distribution of the estimated AFB among prominent AF and non-prominent AF populations constituting the SHHS1 database, using the re-annotated labels.

## 3. Discussion

*Signal quality*

A total of 10.2% of the windows were removed by the signal quality step ($bsqi < 0.8$). This rather high number may be attributable to the poor quality of the signal towards the end of the recordings, when the electrodes are removed, once the patient wakes up, but before the ECG monitor has been switched off by the nurse or the technician. In fact, exclusion of the last 10% of the recordings would reduce window removal to 7.7% (versus 10.2%). In addition, ECG electrodes are often placed in sleep laboratories with less care as compared to Holters or in-hospital 12-lead ECG electrodes, because the ECG is considered to be an auxiliary signal in PSG analysis and not as the primary signal of interest. This further explains the rather large number of windows discarded.

*Error analysis*

After re-annotation, 70 patients were identified as suffering from prominent AF. This represents 2.4% of our population sample. A total of 103 patients were predicted by the RF model to have an AFB exceeding 20%. Two cases of AF were false negatives (FN) i.e. missed by the RF model. In total, there were 35 false positives (FP), i.e., cases inaccurately predicted as prominent AF. This is summarized in Figure 2. Among the FN, one individual presented a highly regular and slow rhythm (40 bpm in average) and was diagnosed as AFL by the intern cardiologists (Figure 5A). The second FN case was classified by the intern cardiologists as PAF. This individual had an estimated AFB of 15% and visual inspection of the windows identified as AF showed that the RF model behavior was accurate (Figure 5B). Thus, this FN resulted from an intrinsic design limitation relating to the 20% decision threshold separating between prominent and non-prominent AF. The false positives (FP) of our model were all due to highly irregular RR interval time series which misled the RF model. Among the 35 FP problematic cases, 14 cases presented a sinus rhythm with high variability (e.g., Figure 6A), 18 presented a sinus rhythm with different forms of ectopic beats (atrial premature complexes and premature ventricular contractions, e.g., Figure 6B), and 3 were diagnosed with other kinds of rhythm (cardiac rhythm exerted by a pacemaker, trigeminy, bigeminy, e.g. Figure 6C).

*Feasibility of diagnosing AF from PSG recordings*

Opportunistic detection of AF using a single and short ECG measurement will miss asymptomatic cases and PAF individuals who are in sinus rhythm during the time of the measurement [4,29]. In this respect, overnight continuous recording and ML driven analysis represents an opportunity to tackle these limitations. In this research, a total of 16 individuals with no documented diagnosis of AF and representing over 22% (16/70) of individuals with prominent AF, were identified by the RF model. These 16 patients had a significantly lower ($p<0.05$, Figure 7) estimated AF burden than the 54 individuals already documented as AF in SHHS1 and confirmed by the intern cardiologists. This stresses that individuals with a lower AF burden are more likely to be undiagnosed and that our approach may identify them. In addition, among the 16 newly identified AF cases, 8 participated in the SHHS2. Among these 8 patients, 4 were documented as AF in SHHS2 and 3 as non-AF. This further strengthen the value of leveraging sleep recordings for the purpose of prominent AF diagnosis.

*Interaction with OSA*

Accurate evaluation of the AFB may enhance phenotyping of OSA, by enabling the identification of individuals whose OSA condition may have already affected cardiac function. Identification of AF was not affected by the presence of OSA. The SHHS1 analysis showed non-inferior PPV for the OSA group compared to the non-OSA group, with OSA defined as AHI≥15 (p=0.03). Among the individuals with prominent AF, 36% (25/70) had an AHI<15 and 64% (45/70) had an AHI≥15. This corresponds to 1.6% of the population with AHI<15 and 3.6% of the population with AHI≥15. Thus, individuals with moderate to severe OSA had significantly (p=0.0004) higher prevalence of AF than the group with AHI<15, which is in accordance with previous studies.

*Newly identified prominent AF cases and strokes*

Among the 16 patients correctly diagnosed with AF and with no previous AF diagnosis documented in the SHHS1, 4 (25%) later had at least one stroke (Table 5). The stroke was fatal for one of these cases (25%). Among these 4 patients, 3 had persistent AF rhythm and 1 had paroxysmal AF rhythm. Three patients had mild OSA ($AHI > 5$) and one severe OSA ($AHI > 30$). In comparison, among the AF cases that were documented in the SHHS1, 13 (19.7%) had at least one stroke. The stroke was fatal in 3 of these cases (23%). Whether the relationship between the phenotype OSA plus AF and stroke is causal or simply associative, these results further highlight the importance in robustly identifying patients with AF in PSG recordings.

*Medicine during sleep: a new perspective*

Today, many digital health home sleep test with portable sensors, have been developed and are already in extensive use for diagnosing sleep disorder breathing. Data-driven algorithms have shown good results at performing automated analysis [30]. This shift to portable digital health monitoring technology has effectively addressed the growing awareness of the number of individuals with sleep disorders. The PSG should be reserved for complicated OSA phenotypes, such as the overlap syndrome between OSA and chronic obstructive pulmonary disease, patients with heart failure or more complicated sleep conditions. These cases should be referred to sleep specialists, whereas obvious cases should be analyzed by other medical professionals, such as general practitioners. Yet, the data collected with PSG or home sleep tests gather valuable physiological measurements that extend beyond the pure traditional sleep diagnosis paradigm. We highlight in this paper the diagnostic potential of data collected during sleep for non-sleep specific conditions, a paradigm we coined earlier as "medicine during sleep" [5]. Although such novel analysis of the data may be fully automated in the distant future, it is likely, in the near future, that the data and its automated analysis will be presented to a medical doctor for review. We thus foresee, a new area of engineering research dedicated to the analysis of data collected during sleep for the purpose of non-sleep-specific diagnosis and monitoring. This also paves the way for a new generation of sleep medical doctors whose training will include a familiarity with the analysis of such data recorded during sleep for diagnosis of non-sleep-specific conditions. To this end, we demonstrated the feasibility of identifying patients with prominent AF from overnight single channel ECG recording.

*Limitations*

A critical limitation of this work was the underlying assumption that a data-driven approach to AF detection can detect patterns in the RR interval time series analysis which are specific to AF. This might be limiting because the model may misclassify patients as AF because they exert a highly irregular RR time series due to other cardiovascular pathologies. This might be an intrinsic limitation of the approach or the expression of a lack of sufficient recordings of cardiac conditions, other than AF, characterized by a highly irregular RR. This is true in particular for patients presenting a high number of ectopic beats. Increasing the representation of such cases in the training process or adding additional discriminative features [31] might improve the ability of the model to learn to distinguish these cases from AF.

A second key limitation was that only some and not all of the recordings in the SHHS1 were reannotated by a medical expert, due to the substantial time it would take for the medical expert to re-annotate all the 2,963 files with a total of 24,400 hours of ECG. Each medical expert spent, on average, 30 minutes reviewing a single overnight ECG recording. Consequently, we may have missed AF cases with no history of AF (as provided by the SHHS1 labels on AF) and with no prediction of a high AF burden. This limitation may be particularly important for the AFL cases which would not manifest with an irregular

RR such as the one case documented in Figure 5A and which was one of the two FN of the RF model. It is also important to note that only a single channel ECG was available from the PSG recordings which is less than in the more traditional 3 or 12-lead Holter ECG used in a classical medical Holter review process. This made the intern cardiologist review process more challenging in distinguishing between noisy sinus waves and fibrillation waves as well as did not enable to confirm the arrhythmias from multiple channels views.

The last main limitation of this study was the AF burden threshold at 20%, used to distinguish between prominent AF and non-prominent AF. This threshold was motivated by the intrinsic capacity of the data-driven algorithm developed and evaluated on the PhysioNet databases, to correctly estimate the AF burden of a patient. Indeed, given the current ability of the RF model, using a smaller threshold would result in many false positives and thus, to too many false alarms for the clinical staff to review. However, training on a larger dataset and with improved features will improve the AF burden estimation and enable lowering the decision threshold on SHHS1.

*Conclusion*

Several pivotal studies conducted over the past decade have highlighted a strong association between obstructive sleep apnea (OSA) and atrial fibrillation (AF). Given the high prevalence of both conditions and their association, there is a high motivation to automatically identify individuals presenting AF in sleep studies in order to better phenotype OSA. We show that individuals with prominent AF can be automatically diagnosed from an overnight single channel ECG recording using digital biomarkers and artificial intelligence and with PPV unaffected by the presence of moderate to severe OSA. This paves the way to novel digital health solution for systematic AF screening in portable sleep studies. We further show that AF detection enabled to identify previously undiagnosed AF cases with over 22% of all AF cases that were previously non-documented. These cases were likely missed by the previous, short AF diagnosis test, but could now be identified by the automated overnight test.

**Financial Disclosure:** none.

**Non-financial Disclosure:** none.

**Data availability:** all four databases used in this research (LTAF, NS, BIHAF, SHHS) are freely available on physionet.org and sleepdata.org

**References**

1. Go, A. S. *et al.* Prevalence of Diagnosed Atrial Fibrillation in Adults. *JAMA* **285**, 2370 (2001).

2. Wolf, P. A., Abbott, R. D. & Kannel, W. B. Atrial fibrillation as an independent risk factor for stroke: The framingham study. *Stroke* **22**, 983–988 (1991).

3. Kirchhof, P. *et al.* 2016 ESC Guidelines for the management of atrial fibrillation developed in collaboration with EACTS. *Kardiol. Pol.* **37**, 2893–2962 (2016).

4. Jones NR, Taylor CJ, Hobbs FR, Bowman L, C. B. Screening for atrial fibrillation: a call for evidence. *Eur. Heart J.* **0**, 1–11 (2019).

5. Behar, J. From sleep medicine to medicine during sleep: a new paradigm. *Sleep* **43**, (2020).

6. Todd, K., McIntyre, W. F. & Baranchuk, A. Obstructive sleep apnea and atrial fibrillation. *Nature and Science of Sleep* (2010).

7. Mehra, R. *et al.* Association of nocturnal arrhythmias with sleep-disordered breathing: The sleep heart health study. *Am. J. Respir. Crit. Care Med.* **173**, 910–916 (2006).

8. Benjafield, A. V. *et al.* Estimation of the global prevalence and burden of obstructive sleep apnoea: a literature-based analysis. *Lancet Respir. Med.* **7**, 687–698 (2019).

9. Boriani, G. *et al.* Detection of new atrial fibrillation in patients with cardiac implanted electronic devices and factors associated with transition to higher device-detected atrial fibrillation burden. *Hear. Rhythm* **15**, 376–383 (2018).

10. Simaityte, M. *et al.* Quantitative Evaluation of Temporal Episode Patterns in Paroxysmal Atrial Fibrillation. *Comput. Cardiol. (2010).* **45**, (2018).

11. Rostagno, C. *et al.* The onset of symptomatic atrial fibrillation and paroxysmal supraventricular tachycardia is characterized by different circadian rhythms. *Am. J. Cardiol.* (1993)

doi:10.1016/0002-9149(93)90454-K.

12. Yamashita, T. *et al.* Circadian variation of paroxysmal atrial fibrillation. *Circulation* (1997) doi:10.1161/01.CIR.96.5.1537.

13. O'Sullivan, J. *et al.* The Accuracy of Smartphone Camera Apps to Detect Atrial Fibrillation: A Systematic Review, Meta-Analysis, Meta-Regression and Modeling Study. *SSRN Electron. J.* (2020) doi:10.2139/ssrn.3478118.

14. Turakhia, M. P. *et al.* Rationale and design of a large-scale, app-based study to identify cardiac arrhythmias using a smartwatch: The Apple Heart Study. *Am. Heart J.* (2019) doi:10.1016/j.ahj.2018.09.002.

15. Perez, M. V. *et al.* Large-scale assessment of a smartwatch to identify atrial fibrillation. *N. Engl. J. Med.* (2019) doi:10.1056/NEJMoa1901183.

16. Dean, D. A. *et al.* Scaling Up Scientific Discovery in Sleep Medicine: The National Sleep Research Resource. *Sleep* **39**, 1151–1164 (2016).

17. Goldberger, A. L. *et al.* PhysioBank, PhysioToolkit, and PhysioNet Components of a New Research Resource for Complex Physiologic Signals. *Circulation* **101**, (2000).

18. Zhang, G. Q. *et al.* The National Sleep Research Resource: Towards a sleep data commons. *J. Am. Med. Informatics Assoc.* **25**, 1351–1358 (2018).

19. Quan, S. *et al.* The Sleep Heart Health Study: Design, Rationale, and Methods. *Sleep* **20**, 1077–1085 (1997).

20. Redline, S. *et al.* Methods for obtaining and analyzing unattended polysomnography data for a multicenter study. Sleep Heart Health Research Group. *Sleep* **21**, 759–767 (1998).

21. Behar, J. A. *et al.* PhysioZoo: A novel open access platform for heart rate variability analysis of

mammalian electrocardiographic data. *Front. Physiol.* **9**, 1–14 (2018).

22. Clifford, G. D., Behar, J., Li, Q. & Rezek, I. Signal quality indices and data fusion for determining clinical acceptability of electrocardiograms. *Physiol Meas* **33**, 1419–33 (2012).

23. Tompkins, W. & Pam, J. A Real-Time QRS Detection Algorithm. *IEEE Trans. Biomed. Eng.* **BME-32**, 230–236 (1985).

24. Behar, J., Oster, J., Li, Q. & Clifford, G. D. ECG signal quality during arrhythmia and its application to false alarm reduction. *IEEE Trans. Biomed. Eng.* **60**, 1660–1666 (2013).

25. Lake, D. E. & Moorman, J. R. Accurate estimation of entropy in very short physiological time series: the problem of atrial fibrillation detection in implanted ventricular devices. *Am. J. Physiol. Circ. Physiol.* **300**, H319–H325 (2010).

26. Sarkar, S., Ritscher, D. & Mehra, R. A Detector for a Chronic Implantable Atrial Tachyarrhythmia Monitor. *IEEE Trans. Biomed. Eng.* **55**, 1219–1224 (2008).

27. Lake, D. E. & Moorman, J. R. Accurate estimation of entropy in very short physiological time series: The problem of atrial fibrillation detection in implanted ventricular devices. *Am. J. Physiol. - Hear. Circ. Physiol.* **300**, (2011).

28. Tunes da Silva, G., Logan, B. R. & Klein, J. P. Methods for Equivalence and Noninferiority Testing. *Biol. Blood Marrow Transplant.* (2009) doi:10.1016/j.bbmt.2008.10.004.

29. Van Gelder, I. C. *et al.* Duration of device-detected subclinical atrial fibrillation and occurrence of stroke in ASSERT. *Eur. Heart J.* (2017) doi:10.1093/eurheartj/ehx042.

30. Behar, J. A. *et al.* Feasibility of Single Channel Oximetry for Mass Screening of Obstructive Sleep Apnea. *EClinicalMedicine* 81–88 (2019) doi:10.1016/j.eclinm.2019.05.015.

31. Carrara, M. *et al.* Heart rate dynamics distinguish among atrial fibrillation, normal sinus rhythm

and sinus rhythm with frequent ectopy. *Physiol. Meas.* **36**, 1873–1888 (2015).

**List of figures**

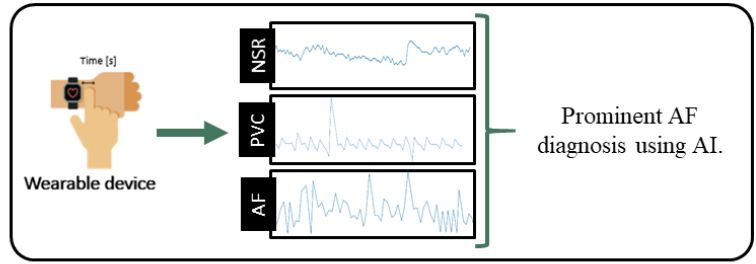
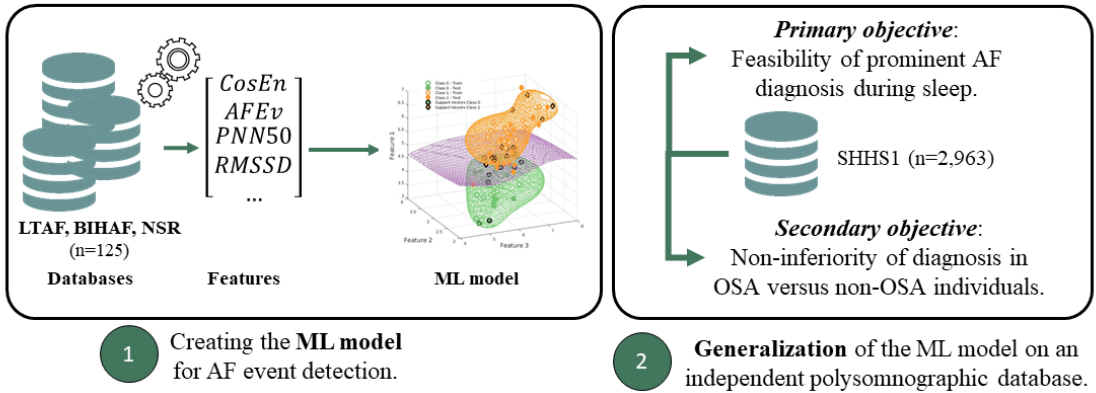

**Figure 1.** Perspective and summary methods of this research. The clinical perspective: systematic screening individuals suspected of sleep-disordered breathing where an important association between AF and obstructive sleep apnea (OSA) has been demonstrated. Summary methods: (1) training databases (LTAF, BIHAF and NSR) are used to train a machine learning (ML) model in recognizing AF events from the beat-to-beat interval time series; (2) the trained model is applied to an independent database of polysomnography recordings (SHHS1) to demonstrate how the classifier generalizes to this independent sleep database and enables to automatically recognize individuals with prominent AF in both non-OSA and OSA individuals.

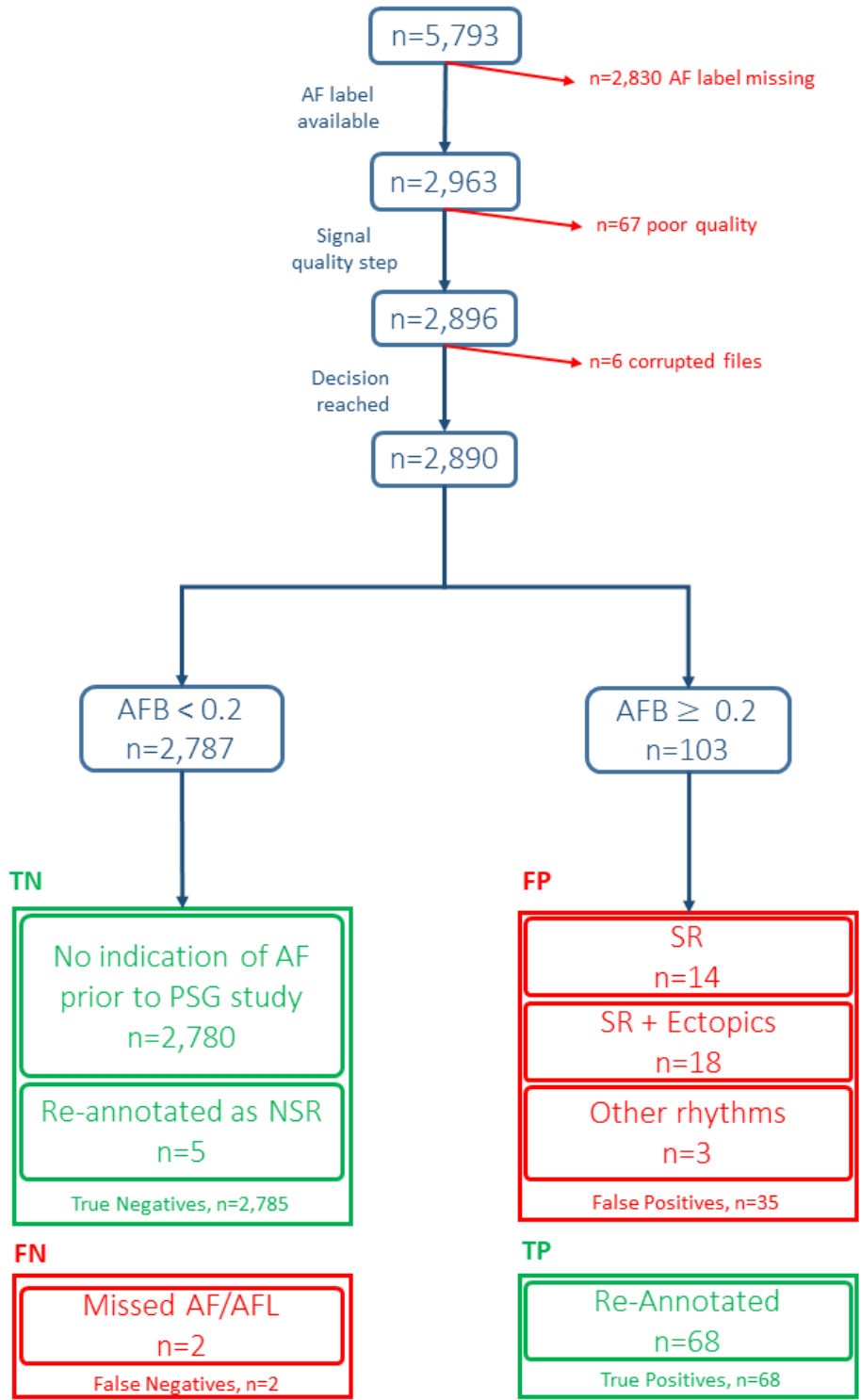

**Figure 2.** Data exclusion process and summary of the per-patient classification results. After the exclusion steps, 2,890 files were considered. Among these, 2,787 were diagnosed with an AF burden (AFB) below 20%, and 103 were diagnosed with an AFB exceeding 20% and hence classified as prominent AF. There were 2 false negative cases with AF missed by the RF model, and 35 false positives flagged as AF despite absence of the condition.

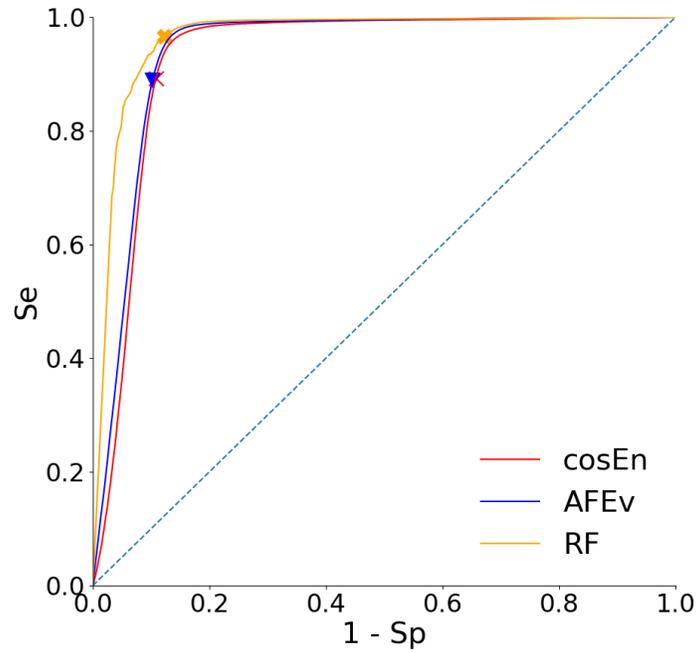

**Figure 3.** ROC curves of the Random Forest classifier compared to two benchmark features widely used in the field of AF detection: the coefficient of Sample Entropy (cosEn, in red), and AFEvidence (AFEv, in blue). This ROC curve is represented for the global model.

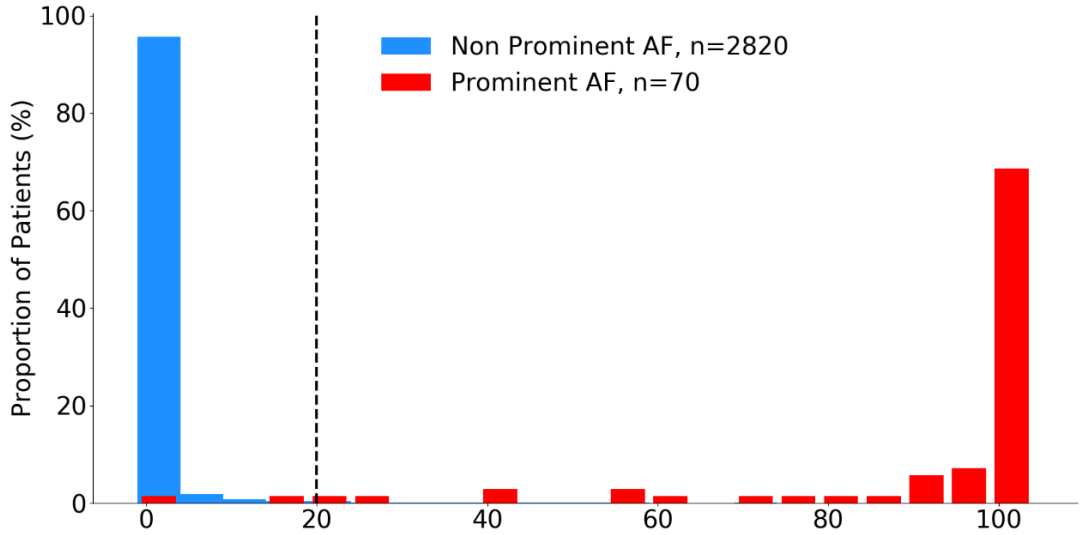

**Figure 4.** Normalized histograms of atrial fibrillation (AF) burden (x-axis) as predicted by the RF model for prominent AF individuals ($AFB \geq 20\%$), determined after the reannotation. The vertical dotted line illustrates the decision threshold used to separate between prominent and non-prominent AF.

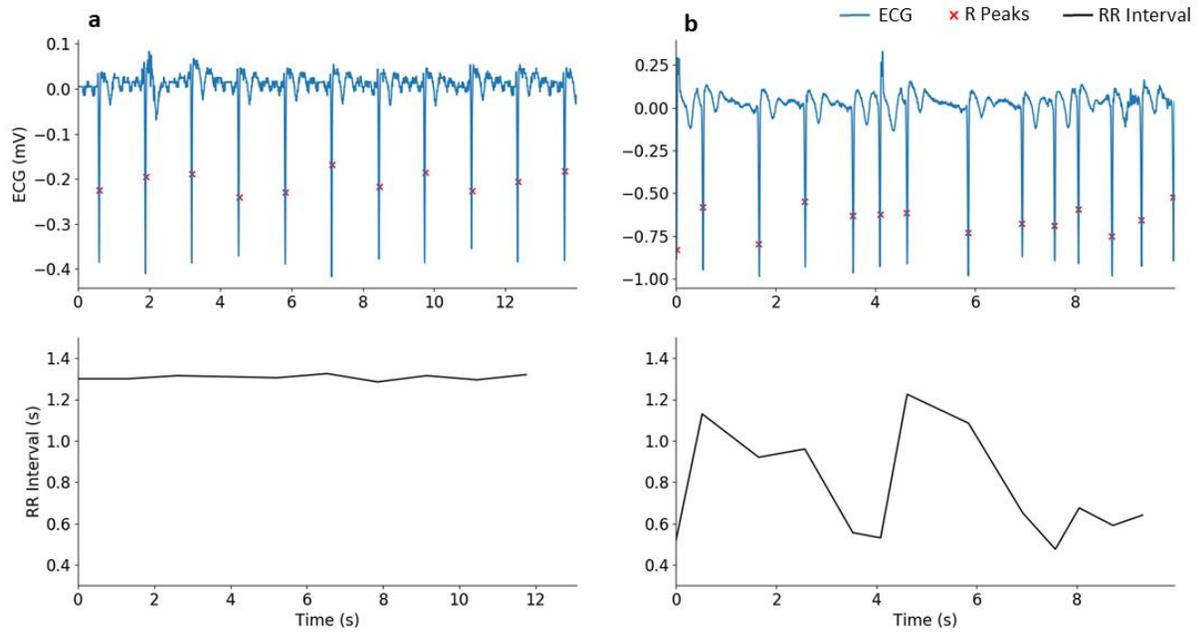

**Figure 5.** ECG recordings and corresponding RR intervals of the two patients missed by the RF model. Panel a: the recording of patient number 202505, diagnosed by the cardiologist as slow atrial flutter (AFL), and missed by the RF model because of his highly regular RR interval. Panel b, the recording of patient number 202845, diagnosed with paroxysmal atrial fibrillation (AF), whose AF events were correctly flagged by the RF model, but presented an AF burden below the threshold (20%) and hence was not flagged as prominent AF.

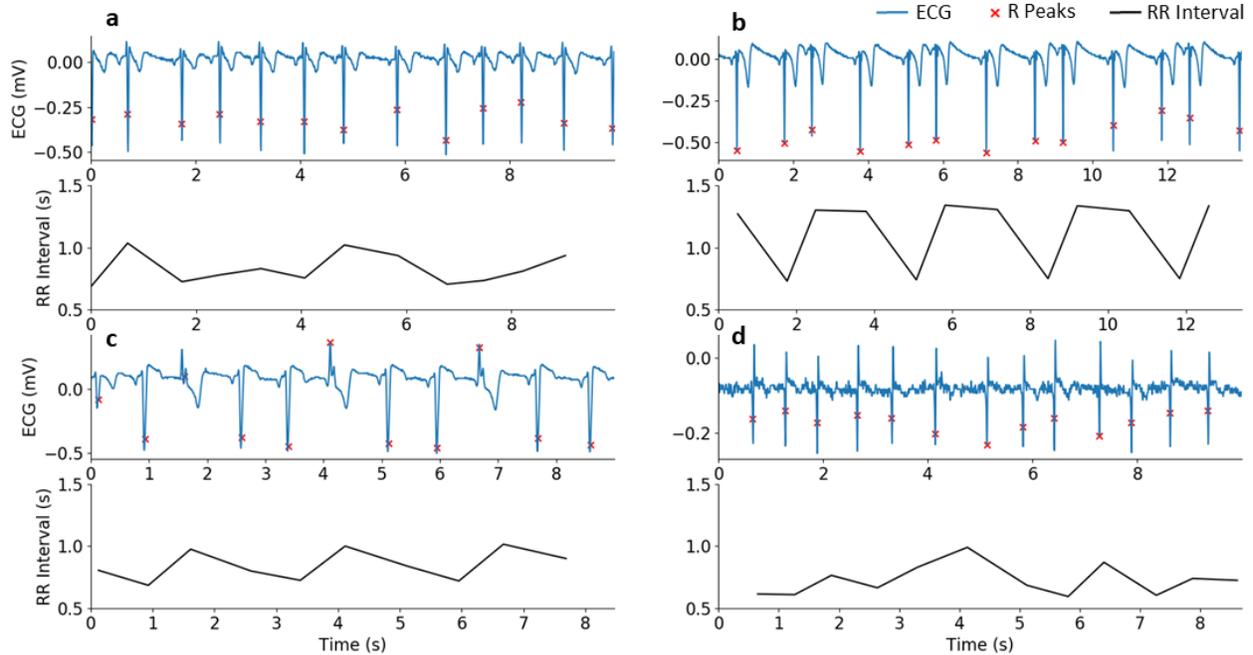

**Figure 6.** ECG recordings of several patients incorrectly classified as AF patients. (a), the recording of patient number 202599 which showed an irregular sinus rhythm; (b), the recording of patient number 202564 which presents a sinus rhythm with atrial premature complexes; (c), recording of patient number 201290 presenting a trigeminy; (d) shows the recording of patient number 201603 which included a noisy RR interval and which could not be classified by the cardiologist.

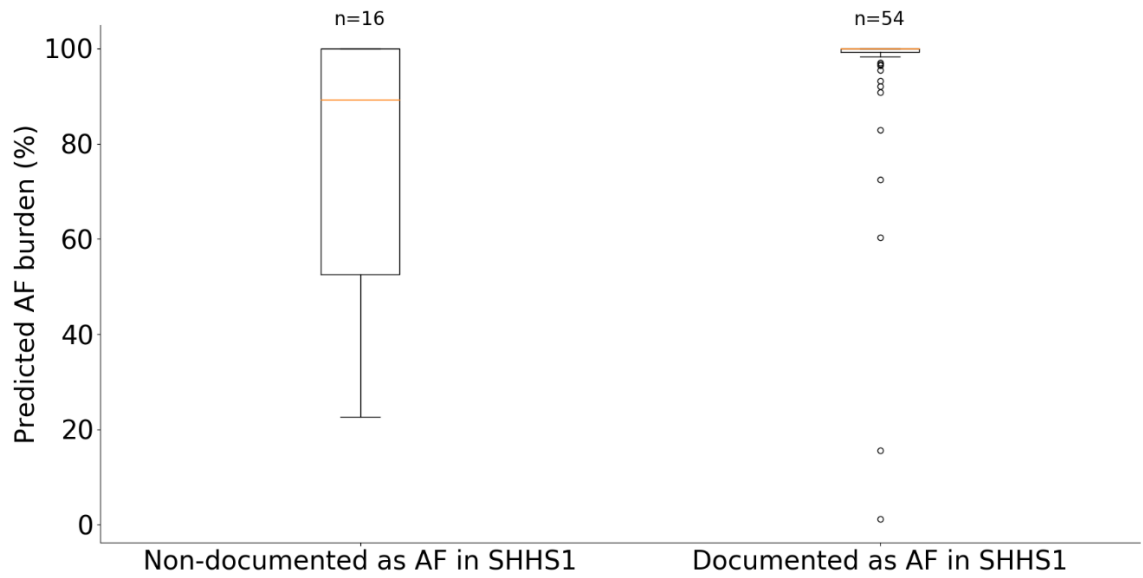

**Figure 7**: Boxplot of the atrial fibrillation burden (AFB) for the individuals already documented in SHHS1 as AF and confirmed by the intern cardiologist and for the cases predicted by our model and confirmed by the intern cardiologist as AF.

**Tables**

**Table 1.** Description of the SHHS1 database for whom AF labels were available (n=2,963). AHI: apnea hypopnea index, ODI: oxygen desaturation index, BMI: body mass index.

|        | **Age** | **AHI** | **ODI** | **BMI** |
|--------|---------|---------|---------|---------|
| Median | 69      | 13.23   | 5.81    | 27.25   |
| Q1     | 53      | 6.53    | 3.38    | 24.55   |
| Q3     | 76      | 23.56   | 10.37   | 30.41   |
| Min    | 39      | 0.0     | 0.0     | 18.0    |
| Max    | 90      | 161.84  | 78.06   | 50.0    |

**Table 2.** List of the features extracted from the RR time-series windows and used within the RF model to distinguish AF events.

| Features | Definition |
| --- | --- |
| **bSQI** | Signal quality of the window [24] |
| **CosEn** | Coefficient of sample entropy [25] |
| **AFE** | AFEvidence [26] |
| **OrC** | Number of points in the bin containing the Origin [26] |
| **IrE** | Irregularity Evidence [26] |
| **PACe** | PAC Evidence [26] |
| **AVNN** | Average NN interval duration |
| **minRR** | Minimum RR interval |
| **medHR** | Median heart rate |

**Table 3.** Performance of the model on the training (NSR, LTAF) and test (BIHAF) sets as well as of the global model (NSR, LTAF and BIHAF).

|        | Training        | Test           | Global Model    |
|--------|-----------------|----------------|-----------------|
| Se     | 0.97            | 0.95           | 0.97            |
|        | (75643/78251)   | (8077/8470)    | (83780/86721)   |
| Sp     | 0.87            | 0.98           | 0.87            |
|        | (89028/102739)  | (10139/10327)  | (99164/113066)  |
| PPV    | 0.85            | 0.98           | 0.86            |
|        | (75643/89423)   | (8077/8265)    | (83780/97682)   |
| NPV    | 0.97            | 0.96           | 0.97            |
|        | (88959/91567)   | (10139/10532)  | (99164/102105)  |
| AUROC  | 0.96            | 0.99           | 0.96            |
| $F_1$  | 0.90            | 0.97           | 0.91            |

**Table 4.** Performance of the model in detecting AF individuals in the SHHS1 database after reannotations of all recordings predicted with $AF\ burden\ \geq\ 0.2$ and review of the positive AF labels previously documented in the SHHS1.

|       | All          | $AHI < 15$   | $AHI \geq 15$ |
|-------|--------------|--------------|---------------|
| Se    | 0.97         | 0.96         | 0.98          |
|       | (68/70)      | (24/25)      | (44/45)       |
| Sp    | 0.99         | 0.99         | 0.99          |
|       | (2785/2820)  | (1561/1579)  | (1224/1241)   |
| PPV   | 0.66         | 0.57         | 0.75          |
|       | (68/103)     | (24/42)      | (44/61)       |
| NPV   | 0.99         | 0.99         | 0.99          |
|       | (2785/2787)  | (1561/1562)  | (1224/1225)   |
| AUROC | 0.99         | 0.99         | 0.99          |
| $F_1$ | 0.79         | 0.72         | 0.83          |

**Table 5**: Summary statistics for newly identified prominent AF individuals who later had one or more strokes.

| ID | Number of strokes | Fatal | AF Burden (%) | Time to stroke (years) | AHI (events/hour) | ODI (events/hour) | Age (years) |
|---|---|---|---|---|---|---|---|
| 202371 | 1 | No | 100 | 0.93 | 60.4 | 49.78 | 77 |
| 204095 | 2 | No | 100 | 1.86 | 7.5 | 4.05 | 80 |
| 202004 | 2 | Yes | 56 | 6.19 | 10.8 | 4.25 | 79 |
| 202462 | 1 | No | 29 | 2.64 | 7.65 | 5.28 | 82 |
| Median (Q1-Q3) | | | 77.6 (48.9-99.8) | 2.25 (1.62-3.52) | 7.7 (7.6-23.1) | 4.76 (4.20-16.40) | 79 (78.5-80) |

**Supplement**

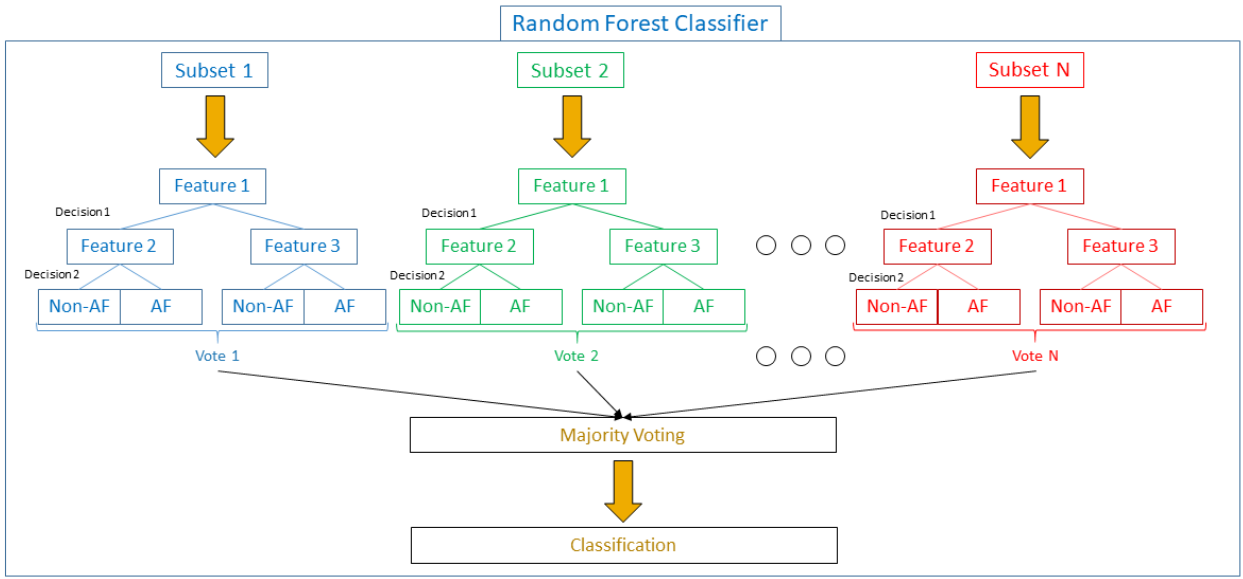

**Supplementary Figure 1.** Illustration of the Random Forest (RF) model used to classify the events as AF or non-AF. The model is fed with a subset of examples containing each nine features derived from beat-to-beat time series. The model decision is based on a subset of decision trees with a given maximal depth (here, maximal depth of 2). Each tree is trained based on a subset of the original training set drawn with replacement. Eventually, the model classifies a given example based on the majority vote among the trees.